\documentclass[8.5pt,twoside,twocolumn]{article}
\usepackage[super,sort&compress,comma]{natbib} 
\usepackage{mhchem}
\usepackage{times,mathptmx}
\usepackage{sectsty}
\usepackage{balance} 

\usepackage{graphicx} 
\usepackage{lastpage}
\usepackage[format=plain,justification=raggedright,singlelinecheck=false,font=small,labelfont=bf,labelsep=space]{caption} 
\usepackage{fancyhdr}
\usepackage{bm}
\usepackage{comment}
\usepackage{amsmath,amssymb,latexsym}

\newcommand{\NL}{N_\mathrm{L}}
\newcommand{\NK}{N_\mathrm{K}}
\newcommand{\rhoL}{\rho_\mathrm{L}}
\newcommand{\rhoK}{\rho_\mathrm{K}}
\newcommand{\rhoLK}{\rho_\mathrm{LK}}

\pdfoutput=1

\newcommand{\figurewidth}{0.46\textwidth}

\begin{document}

\thispagestyle{plain}
\fancypagestyle{plain}{

\renewcommand{\headrulewidth}{1pt}}
\renewcommand{\thefootnote}{\fnsymbol{footnote}}
\renewcommand\footnoterule{\vspace*{1pt}%
\hrule width 3.4in height 0.4pt \vspace*{5pt}} 
\setcounter{secnumdepth}{5}

\makeatletter 
\def\subsubsection{\@startsection{subsubsection}{3}{10pt}{-1.25ex plus -1ex minus -.1ex}{0ex plus 0ex}{\normalsize\bf}} 
\def\paragraph{\@startsection{paragraph}{4}{10pt}{-1.25ex plus -1ex minus -.1ex}{0ex plus 0ex}{\normalsize\textit}} 
\renewcommand\@biblabel[1]{#1}            
\renewcommand\@makefntext[1]%
{\noindent\makebox[0pt][r]{\@thefnmark\,}#1}
\makeatother 
\renewcommand{\figurename}{\small{Fig.}~}
\sectionfont{\large}
\subsectionfont{\normalsize}

\renewcommand{\headrulewidth}{1pt} 
\renewcommand{\footrulewidth}{1pt}
\setlength{\arrayrulewidth}{1pt}
\setlength{\columnsep}{6.5mm}
\setlength\bibsep{1pt}

\twocolumn[
  \begin{@twocolumnfalse}
\noindent\LARGE{\textbf{Self-assembly of colloidal polymers via depletion-mediated lock and key binding$^\dag$}}
\vspace{0.6cm}

\noindent\large{\textbf{Douglas J. Ashton,$^{\ast}$\textit{$^{a}$} Robert L. Jack,\textit{$^{a}$} and
Nigel B. Wilding\textit{$^{a}$}}}\vspace{0.5cm}

\noindent\textit{\small{\textbf{Received Xth XXXXXXXXXX 20XX, Accepted Xth XXXXXXXXX 20XX\newline
First published on the web Xth XXXXXXXXXX 200X}}}

\noindent \textbf{\small{DOI: 10.1039/b000000x}}
\vspace{0.6cm}

\noindent \normalsize{We study the depletion-induced self-assembly of indented colloids.
Using state-of-the-art Monte Carlo simulation techniques that treat
the depletant particles explicitly, we demonstrate that colloids assemble by a
lock-and-key mechanism, leading to colloidal polymerization. 
The morphology
of the chains that are formed depends sensitively on the size of the
colloidal indentation, with smaller values additionally permitting
chain branching.  In contrast to the case of spheres with attractive
patches, Wertheim's thermodynamic perturbation theory fails to provide a fully
quantitative description of the polymerization transition.  We trace
this failure to a neglect of packing effects and we introduce a modified theory
that accounts better for the shape of the colloids, yielding improved
agreement with simulation.
}

\vspace{0.5cm}
 \end{@twocolumnfalse}
  ]

\newcommand{\vfs}{\eta_s^r}

\section{Introduction} \label{sec:intro}

The goal of self-assembly is to tailor the interactions among
nano-scale particles so that they spontaneously assemble themselves
into functional materials or
devices~\cite{whitesides:2002,Glotzer2007}.  Such processes are
widespread in biology, where they have been optimised by evolution so
that assembly is rapid and reliable.  However, mimicking this
behaviour in the laboratory involves many challenges, particularly the
design and synthesis of particles whose interactions can be accurately
predicted and controlled.  Notable experimental successes have
included assembly of unusual crystals from either ``patchy'' or
DNA-functionalised colloids~\cite{Nykpanchuk08,Chen11}.  More
recently, particles have been shown to self-assemble into structures
that depend strongly on their \emph{geometrical
shapes}~\cite{Sacanna:2010ys,marechal2010,Rossi11,Henzie12,octapod12,sacanna13, Kraft12},
and the role of shape and packing effects in self-assembly has also
attracted theoretical and computational
interest~\cite{Damasceno12,Odriozola08-13,Torquato09,Henzie12,octapod12,Kraft12,marechal2010,
Ni12,Damasceno12-nano,Haji13,Gantapara13,Anders-arxiv13}.
Here, we use computer simulations to show how self-assembly of
indented colloidal particles can be tuned by subtly varying their
shape and interactions, in a manner that should be accessible in
experiment~\cite{Sacanna:2010ys}.

\begin{figure} 
  \includegraphics[width=\figurewidth]{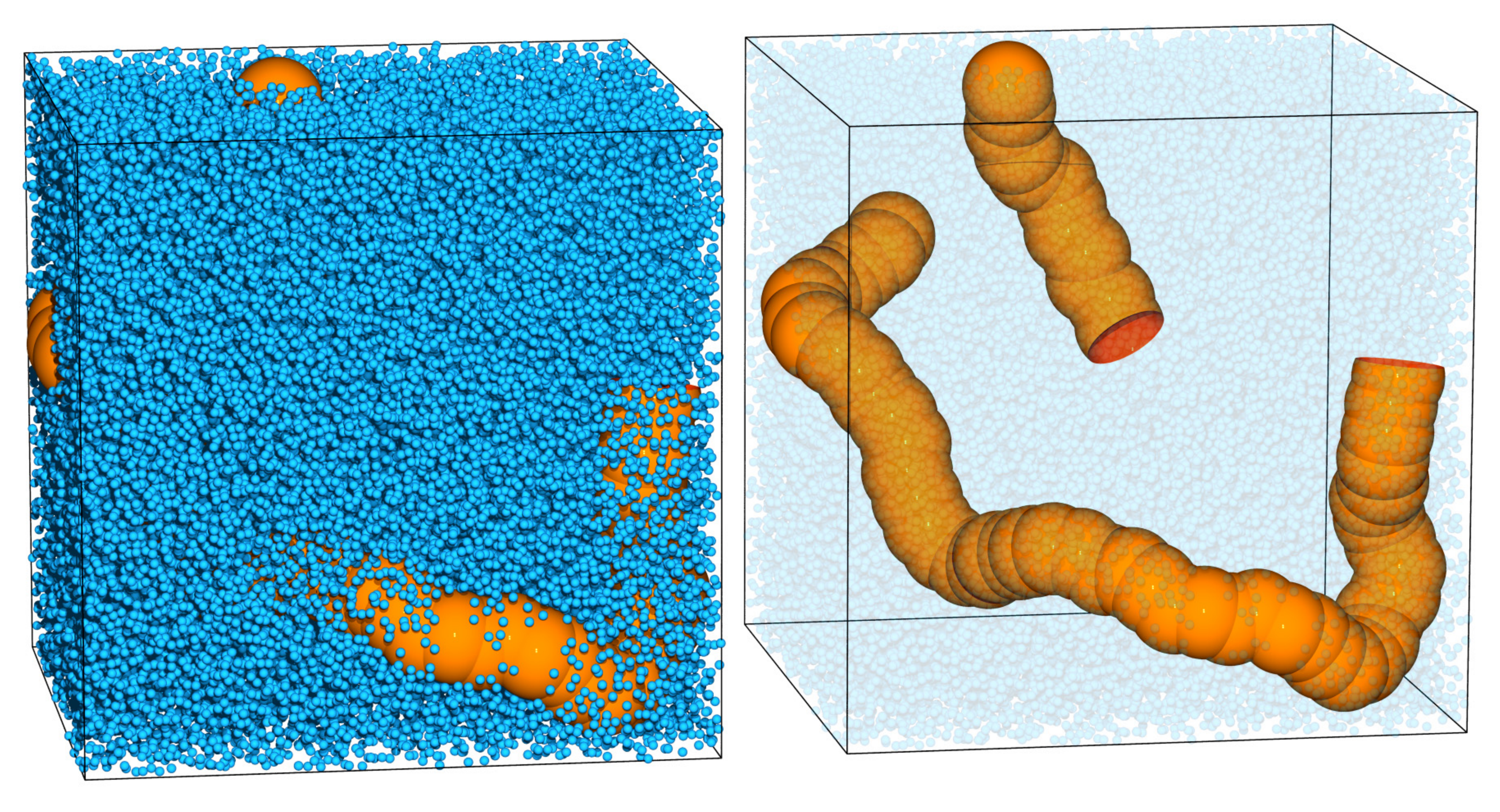}
  \caption{Equilibrium snapshot of self-assembled chains of spherically-indented colloids.  
  The depletant is shown explicitly in the left panel but suppressed in the right panel, for clarity.
  System parameters are $h=0.7$ and $\eta^{\rm r}_s=0.094$: see text for details. }
  \label{fig:snapshot0.3}
\end{figure}

To this end, we exploit {\em depletion
  forces}~\cite{Lekkerkerker:2011}, which enable the precise control
of particle interactions that is required for self-assembly.
Depletion is an attractive interaction between colloid particles that
arises when they are mixed with much smaller `depletant' particles,
for example polymers or another species of colloid.  These forces are
particularly strong in colloids with complementary geometrical forms,
such as buckled spheres~\cite{Sacanna:2010ys} or bowl
shapes~\cite{marechal2010}.  Such systems can assemble via ``lock and
key binding'' in which the convex part of one particle interlocks with
the concave part of
another~\cite{Konig2008,Sacanna:2010ys,Odriozola08-13}.
Fig.~\ref{fig:snapshot0.3} shows the results of a computer simulation,
where colloids with self-complementary shapes~\cite{Sacanna:2010ys} have assembled
themselves into chains, in the presence of depletant particles.  We
show in the following that the properties of these ``colloidal
polymers'' can be controlled through the colloidal shape and the
depletant number density.  The persistence length of the polymer
depends on the colloidal shape, and for some shapes, the chains can
also branch, leading to interconnected networks of particles.

We emphasize that the complementary shapes of colloidal particles~\cite{Sacanna:2010ys,Sacanna2011,Bahadur2012} and
properties of the depletion interaction can both be measured and controlled
in experiments.  Indeed, some depletants even allow colloid
interactions to be tuned \emph{in
situ}~\cite{Alsayed04,Savage09,Rossi11}, potentially leading to
real-time adaptive control of interaction
parameters~\cite{Klotsa:2013vn}.  However, the experimental parameter
space associated with mixtures of colloid and depletant particles is
very large, depending on the size, shape and concentration of both
species.  Theory and computer simulation can therefore offer
guidance for experiment, by predicting the parameters for which robust
assembly occurs, and the likely nature of the self-assembled products.
We argue that such simulations should deal explicitly with the
depletant particles, both in the interests of reproducing the
experimental reality and for avoiding the need to develop effective
(``depletion'') potentials, which for irregularly shaped particles
represents a formidable task.

We give details of our model and simulation techniques in Sec.~\ref{sec:simdetails},
with numerical results in Sec.~\ref{sec:results}.  In Sec.~\ref{sec:theory} we discuss
how Wertheim's theory of associating fluids can be applied to this system.  Our
conclusions are summarised in Sec.~\ref{sec:conclusions}.  In addition, some further
details of our theoretical calculations are presented in electronic supplementary information.

\section{Model and simulation methods} \label{sec:simdetails}

We used state-of-the-art Monte Carlo (MC) simulation
techniques to study spherically-indented colloids, together with
smaller hard sphere particles which act as a depletant.  The shape of
each indented colloid begins as a hard sphere of diameter
$\sigma_l=1$, from which an indentation is formed by cutting away a
sphere of the same diameter, whose center is a distance $d_c$ from the
center of the original sphere.  Thus, the dimensionless depth of the
indentation is $h\equiv 1 - d_c/\sigma_l$. Our systems contain $N=60$
colloid particles in a box of size $V$, with a number density
$\rho=N/V=0.2\sigma_l^{-3}$, and we consider values of $h$ between
$0.3$ and $0.7$.  The hard spheres comprising the depletant fluid have
diameter $\sigma_s=0.1\sigma_l$.  The colloid shape and the size ratio
between colloids and depletant are consistent with experimental
studies~\cite{Sacanna:2010ys}.

\begin{figure} 
\includegraphics[width=\figurewidth]{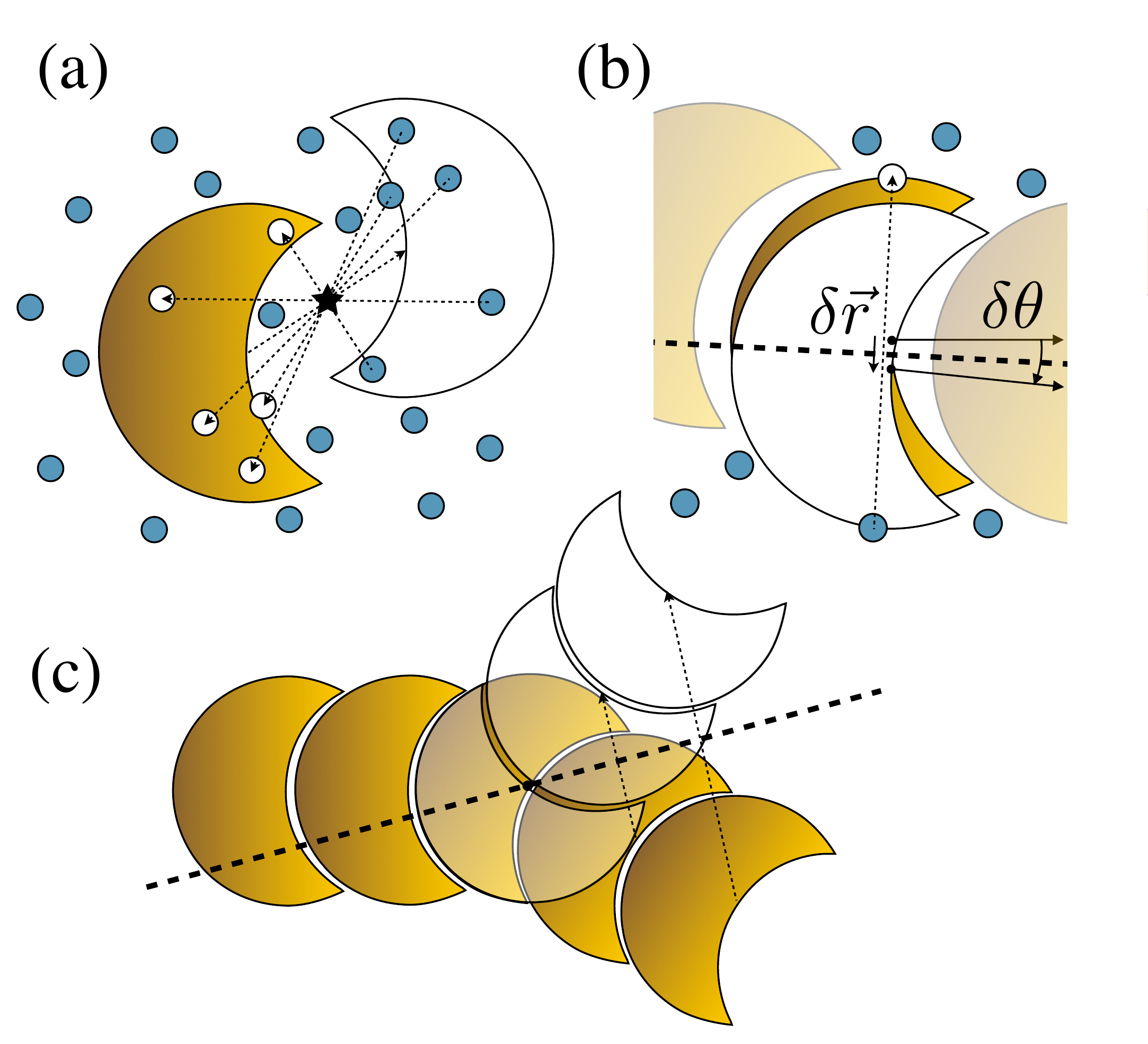}
  \caption{GCA move set for indented particles. {\bf (a)}~A
    colloid reflects through a point pivot to a new position
    (outlined). Any particles that overlap in the new position reflect
    through the same pivot to occupy the space vacated by the
    colloid. {\bf (b)} Constrained plane reflections  (see text) 
    allow for small scale vibrations of colloids within a chain. {\bf (c)}~Reflecting a colloid
    in a plane passing through its center moves it to a new position (outlined). When overlapping particles are similarly reflected, the
     chain `flexes'.}
\label{fig:GCA}
\end{figure}

To obtain accurate computational results for this system, we use a
variant of the geometrical cluster algorithm (GCA) \cite{Liu2004}.
This is a sophisticated Monte Carlo scheme that 
 updates large groups (``clusters'') of
particles, with both colloids and depletants moving together.  
The scheme respects detailed balance, ensuring that it samples the
Boltzmann distribution of the system.  Use of
such a specialized technique is essential for coping with the
disparity in size between colloids and depletant.  Standard Monte
Carlo and Molecular Dynamics techniques are unequal to the task of
relaxing such systems because the depletant acts to frustrate
colloidal motion except on very small length scales. This problem can
be readily appreciated from Fig.~\ref{fig:snapshot0.3}.


The GCA is based on self-inverse geometric operations that can be
tailored to effectively sample the system of interest.  In the case of
self-assembled structures it is essential that relaxation occurs on
all length scales to ensure ergodicity. To this end we use the
combination of updates described in Fig.~\ref{fig:GCA}: A ``pivot''
(point reflection) operation (see Fig.~\ref{fig:GCA}(a)) is employed to
relax particle positions, while a plane reflection
operation~\cite{sinkovits:144111} allows colloids to sample different
orientations.  We combine these two kinds of move to sample the equilibrium
state of the system.  

Pivot moves 
are  effective in
moving clusters of colloids that have started to assemble.
These moves are rejection-free by construction, and the pivot
point is chosen at random.
For reflection moves, we choose the reflection planes to
aid relaxation of single monomers within their binding pockets, and to
promote flexing of the colloidal chains.  For the former case, we use moves where 
the reflection plane is constrained to lie close to the orientation vector of a monomer in the
chain (Fig.~\ref{fig:GCA}(b)). In the latter case, the 
plane is placed through the center of the monomer,
at an arbitrary angle to the orientation vector (Fig.~\ref{fig:GCA}(c)). 
Since the reflection plane is not typically placed along
one of the box axes, care must be taken as the cluster move may
conflict with the periodic boundary conditions. We avoid this problem by rejecting moves in which
any particle in a cluster is interacting with a particle from another
periodic image of the system~\cite{sinkovits:144111}.
All
updates exploit a highly efficient hierarchical overlap search
algorithm that allows us to determine whether a proposed move leads to
overlaps of our anisotropic particles
\cite{Marechal:2010fk,He:1990kx,Cinacchi:2010uq}.

The depletant particles are treated grand-canonically in our simulations.
That is, their number is free to fluctuate, corresponding to the
common experimental situation of a depletant that is in equilibrium
with a bulk reservoir.  We therefore quote the depletant reservoir
volume fraction $\vfs=N_s\pi\sigma_s^3/(6V)$ as a measure of the
driving force for depletion induced assembly, where $N_s$ is the
average number of depletant particles.

\section{Results} \label{sec:results}

\begin{figure} 
  \includegraphics[width=\figurewidth]{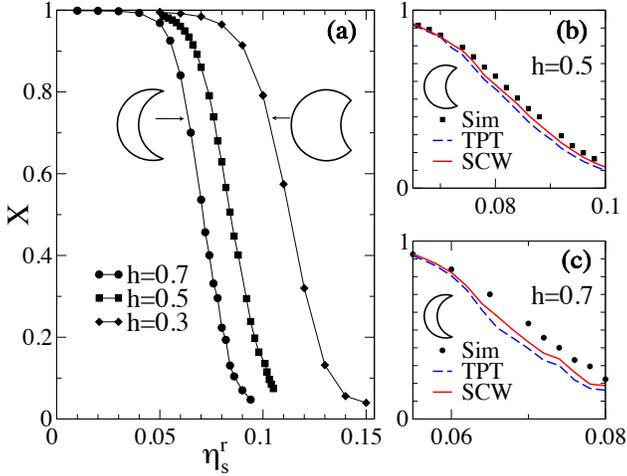}
\caption{{\bf (a)} Simulation estimates of $X(\eta_s^r)$ for
 $h=0.3,0.5,0.7$; lines are guides to the eye and uncertainties are
 comparable to the symbol sizes. Comparison of the simulation data (Sim)
 with the predictions of TPT and SCW (see text) in the non-branching regime {\bf (b)} $h=0.7$, and {\bf (c)} $h=0.5$.
 There are numerical uncertainties in the TPT/SCW predictions which are comparable with those in
 the simulation data: these arise from the numerical estimation
 of $f_{\rm A}$ from simulations containing two colloids with depletant.}
  \label{fig:X}
\end{figure}

We now present our simulation results. We first assess how the degree of
polymerization depends on the depletant volume fraction $\eta_s^r$.
To this end, we label the indentation on each colloidal particle as
its ``lock site''. The concave part of the surface acts as the
``key'', which fits snugly in the lock.  Let $\NL$ be the average number of
lock sites that are available for binding (where no other colloidal
particle is already bound), and let $\NK$ be the average number of colloidal
particles that are not currently occupying any lock site.
(Occupation
of a lock site is decided on the basis of a radial cutoff criterion;
results are insensitive to the choice of this cutoff).
The 
number densities of such particles are then
$\rhoL=\NL/V$ and $\rhoK=\NK/V$.
We also define
$X=\rhoL/\rho$.  For an unassociated fluid, $X\approx 1$; for a system
consisting of long colloidal polymers then $X\approx 0$.  If the
polymers are `tree-like' (without closed loops), then the average
degree of polymerization is $1/X$.

\begin{figure} 
  \includegraphics[width=\figurewidth]{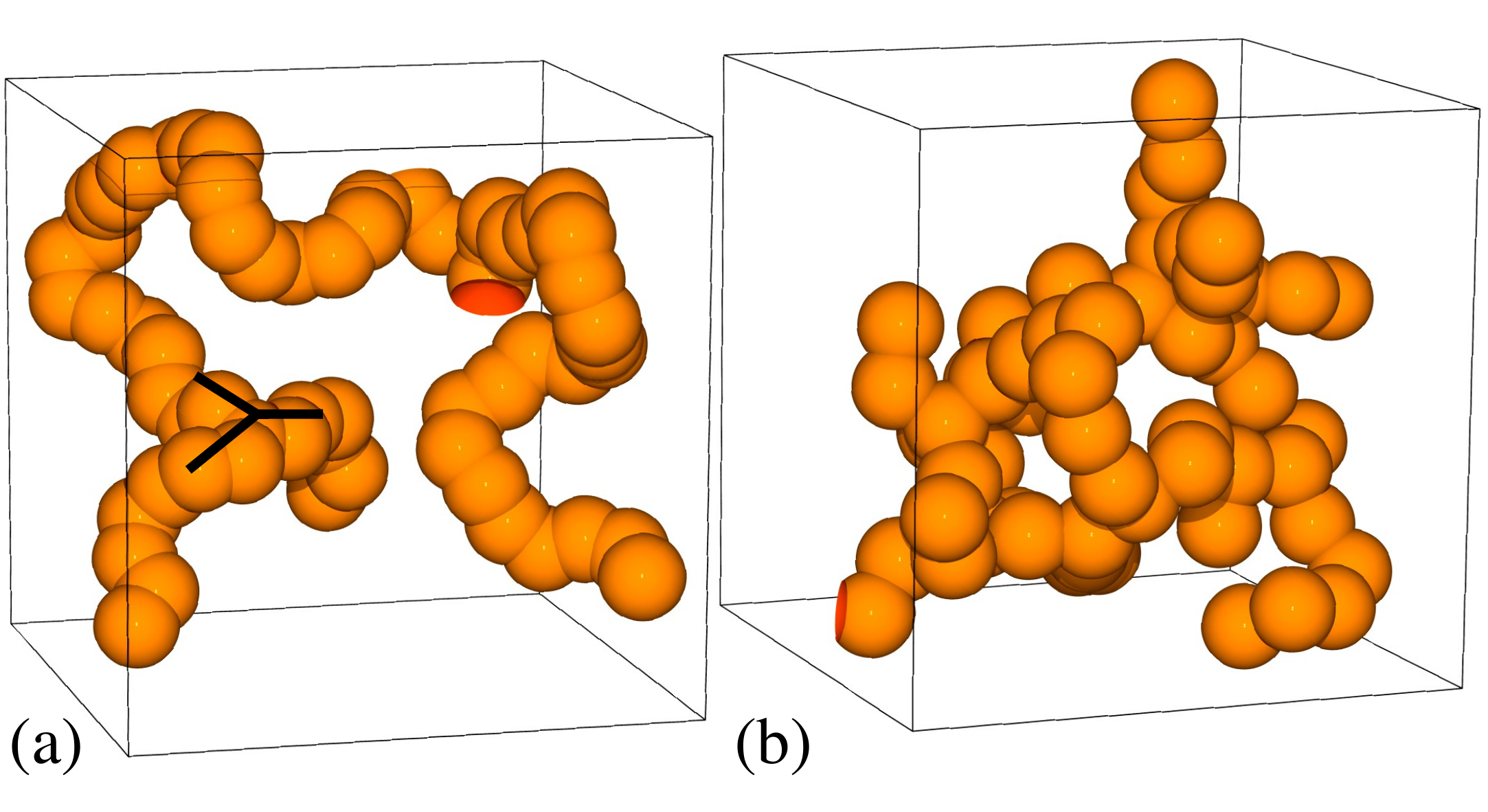}

\caption{Snapshots for $X<0.1$ (depletant not shown) and varying
    indentation depth. {\bf (a)} At $h=0.5$, $\eta_s^r=0.105$, this
    system consists of just two large chains. A single branch point is
    also indicated. {\bf (b)} At
    $h=0.3$, $\eta_s^r=0.14$, the polymers form an
    interconnected network of chains~\cite{Bianchi06}.}
  \label{fig:other-snaps}

\end{figure}

Fig.~\ref{fig:X}(a) depicts our measurements of $X$ (black circles) for
various $h$, as the depletant volume fraction $\eta_s^r$ is
increased. The range of $\vfs$ over which polymerization occurs is
quite narrow in each case (particularly for deep indentations), and
this range is shifted to smaller $\vfs$ as $h$ increases.  Physically,
the lock-and-key binding is strongest when the colloid indentations
are deep, and the shape complementarity is most pronounced.  At the
largest values of $\eta_s$, almost all the colloids are members of
chains, $X\approx 0$.  As well as the results shown here for $N=60$ colloidal particles, we have also performed a limited set of simulations for $N=120$, under the same conditions. We find fully quantitative agreement between results for these two system sizes, indicating that finite-size effects are relatively small, at least for the quantities measured here

Fig.~\ref{fig:other-snaps} shows snapshots of the equilibrated polymer
configurations that form for $h=0.5$ and $h=0.3$, at values of
$\eta_s^r$ corresponding to $X\approx 0.1$. Compared with the results
for $h=0.7$ (Fig.~\ref{fig:snapshot0.3}) one observes that deeper
indentations result in stiffer chain conformations. To quantify these
differences, we have measured the persistence length $b$, defined
through $\langle \cos(\theta_k)\rangle=e^{-k\sigma_l/b}$, where $\theta_k$ is
the angle between orientation vectors of colloid particles that are
$k$th neighbours in the chain. Thus, large values of $b$ correspond to
stiff chains: for $h=(0.3,0.5,0.7)$, we find
$b=(1.0,3.3,9.1)\sigma_l$.

\begin{figure} 
\includegraphics[width=6cm]{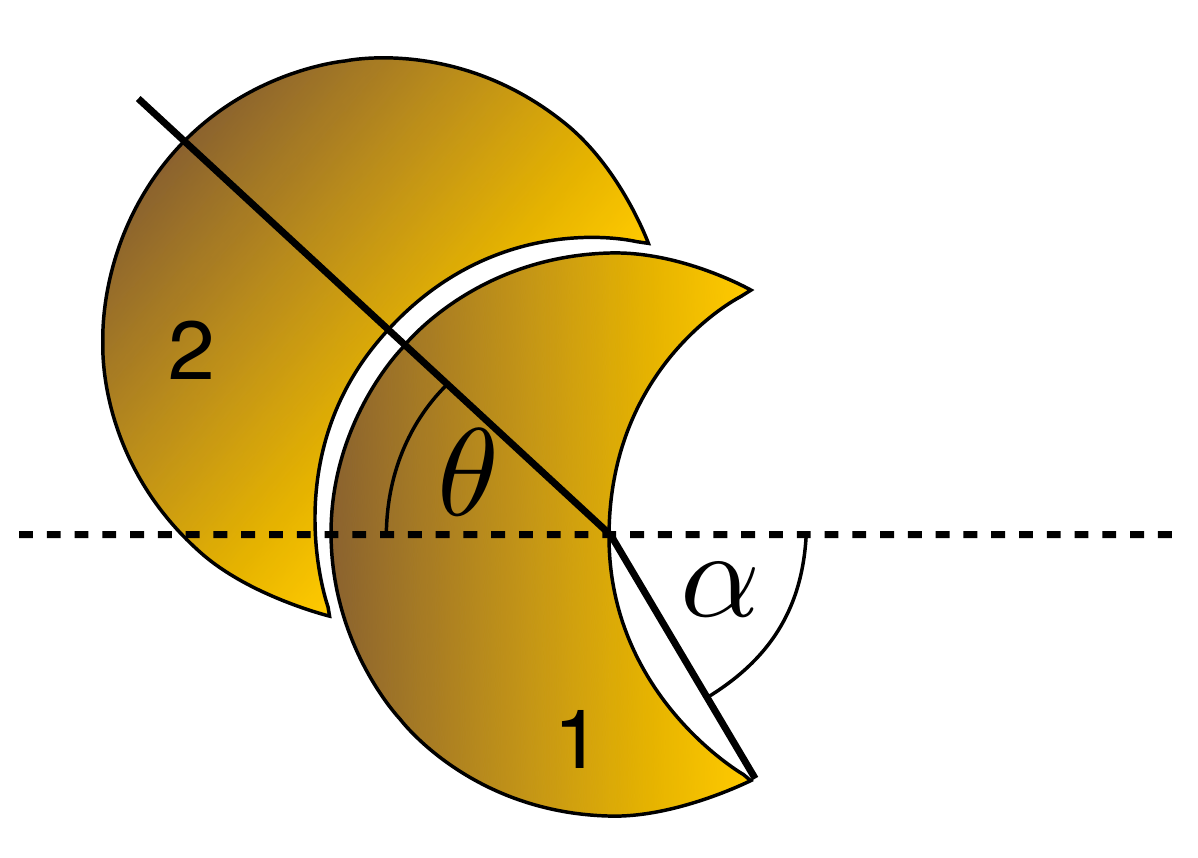}
  \caption{Two indented particles (labelled 1 and 2) that form part of a colloidal polymer.
  The condition that the concave surface of particle $2$ is entirely in contact with the surface
  of particle $1$ is $\theta < \pi-2\alpha$.  From the definition of $h$, we have $\cos\alpha = 1-h$.
  }
  \label{fig:persist}
\end{figure}

\vspace{6pt}

These values can be explained by 
a simple geometrical argument, illustrated by the two particles shown in
in Fig.~\ref{fig:persist}.  The angle $\alpha$ depends on the colloidal shape, as $\cos\alpha = (1-h)$.
 We suppose that particle $2$ can bind on any part of the surface of 
lock $1$, as long as $\theta < \pi-2\alpha$, where $\theta$ is the angle between the orientation vectors 
of the colloids as shown.  For particles in contact, this is the condition that all of the concave
surface of particle $2$ is in contact with the convex surface of particle $1$.
For bonded particles, we recognise $\theta$ as the usual polar angle
in spherical co-ordinates.  To calculate the average of this quantity
subject to the constraint $\theta<\pi-2\alpha$, we write
  $\langle \cos \theta \rangle = (1/Z) \int_0^{\pi-2\alpha} \cos\theta \sin\theta d\theta$
where 
  $Z = \int_0^{\pi-2\alpha} \sin\theta d\theta$ is a normalisation constant
(the volume element $\sin\theta d\theta$ arises from the spherical geometry, as usual).  
Evaluating the integrals yields $\langle\cos\theta\rangle = \frac12[1+\cos(\pi - 2\alpha)]$.
%
From the definition of $\alpha$ one has $\cos(\pi-2\alpha) = 1-2(1-h)^2$, and assuming that angles 
along a chain are independently distributed in this way,
one has ${\rm e}^{-\sigma_l/b} = \langle \cos\theta \rangle $, yielding the relation $\sigma_l/b = -\log[h(2-h)]$
For the colloids with
$h=(0.3,0.5,0.7)$ considered here, this argument predicts $b\approx(1.5,3.5,10.6)\sigma_l$, 
in reasonable agreement with the simulation result given above.
These results illustrate how the
properties of self-assembled colloidal polymers may be controlled
through the geometrical shape of the colloids.

For shallower indentations the assembled polymers may support
branching.  This is only possible when the indentation is 
small enough for two colloid particles to
``lock
onto'' the convex part of a third one.  The marginal case is
$h=0.5$, for which a single key surface can just accommodate
two locks.  As $h$ decreases, the branching probability increases
rapidly.  When bonds are strong ($X<0.1$ as in
Fig.~\ref{fig:other-snaps}), we find that the fractions of particles
involved in branching for $h=(0.5,0.4,0.3)$ are $(1\%, 9\%,15\%)$.
Again, by changing the colloid shape, the self-assembled chains can be
varied from linear polymers ($h=0.7$) to chains with a few
branches ($h=0.5$), and finally ($h=0.3$) to strongly
branched structures.  In the strongly-branched case, we also sometimes find a cluster of bound particles that 
percolates (spans the simulation box).  More detailed characterisation of both percolation
transitions and liquid-vapour phase transitions in this system would be useful avenues for
future study, but they are beyond the scope of this work, due to the computational difficulty
associated with our exact treatment of the depletant fluid.

\section{Theory} \label{sec:theory}

Given the range of chain lengths, persistence lengths and branch-point densities that are
 possible on varying just the
depletant density $\vfs$ and the indentation depth $h$, theoretical
insight is very valuable in guiding choices of colloidal geometry and
depletant parameters, both in simulation and, potentially, in
experiment.  We have applied Wertheim's theory of associating
fluids~\cite{Wertheim} to these indented colloids, following the work
of Sciortino and co-workers~\cite{Bianchi06,Sciortino2007} on `patchy'
colloids.  This theory generalizes liquid state theory, incorporating
steric constraints.  For example,
at most one particle may occupy any lock site; we also assume that
chain branching may not occur, which is valid for $h\gtrsim 0.5$. 
Within the theory, depletion interactions appear as two-body effective interactions
between the colloidal particles, obtained formally by integrating out
the depletant fluid.  Based on these assumptions, Wertheim's theory
gives a diagrammatic series for the density functional of the system,
from which the number densities $\rhoL$ and $\rhoK$ may be derived.
This section contains a summary of this theoretical analysis, concentrating on the physical
insight it provides.  In the supplementary information~\cite{SI}, we
provide the formulae that we use to obtain the predictions in Fig.~\ref{fig:X}, although 
we defer the proofs of these formulae to a later publication.

At leading order, Wertheim's theory 
reduces to the familiar law of mass action: $\rhoLK = \rhoL \rhoK K_0$ where $\rhoLK = \rho - \rhoL$ is the number
density of bonds (i.e. the number density of occupied lock sites), and
$ K_0 $ is the bare 
equilibrium constant, which depends on the attractive forces between
particles.  
%
The law of mass action applies quite accurately in the dilute limit $\rho\sigma_l^3\ll 1$, but
to go beyond this limit, one must also take account of
repulsive forces between particles, and the resulting packing
effects. The second part of Wertheim's theory achieves this by a perturbative
expansion about a reference system without any attractive interactions ($\vfs=0$).  
The theory is therefore accurate if the packing of particles in the
presence of attractive forces is very similar to their packing in the
reference system.  Formally, the thermodynamic perturbation theory
(TPT) of Wertheim approximates the density functional of the system by an
infinite subset of terms in its diagrammatic expansion.  The result is
that the bare equilibrium constant $ K_0 $ in the law of mass action is replaced by~\cite{Sciortino2007} 
%
\begin{equation}
K = \frac{1}{\Omega} 
\int \mathrm{d}\bm{r}_{12} \, \mathrm{d}\omega_2 \,  g_{\rm R}(\bm{r}_{12},\omega_1,\omega_2) 
f_{\rm A}(\bm{r}_{12},\omega_1,\omega_2)  .
\label{eq:TPT_K}
\end{equation}
Here $\omega_1,\omega_2$ represent the orientations of two particles, with $\bm{r}_{12}$ the vector between them,
and $\Omega$ is the phase space volume associated with one particle's orientation.
Also, $g_{\rm R}(\bm{r},\omega_1,\omega_2)$ is the two-particle distribution function in the reference
system (without attractions), while $f_{\rm A}(\bm{r},\omega_1,\omega_2)$ is a Mayer-$f$ function associated with
the attractive part of the effective interactions between particles.

In the dilute limit, $g_{\rm R}=1$ except when the two particles overlap, and
one recovers the standard formula for the bare equilibrium constant $K=K_0$.  Outside the dilute regime,
particle packing effects are taken account of through $g_{\rm R}$.

\begin{figure} 
  \includegraphics[width=7cm]{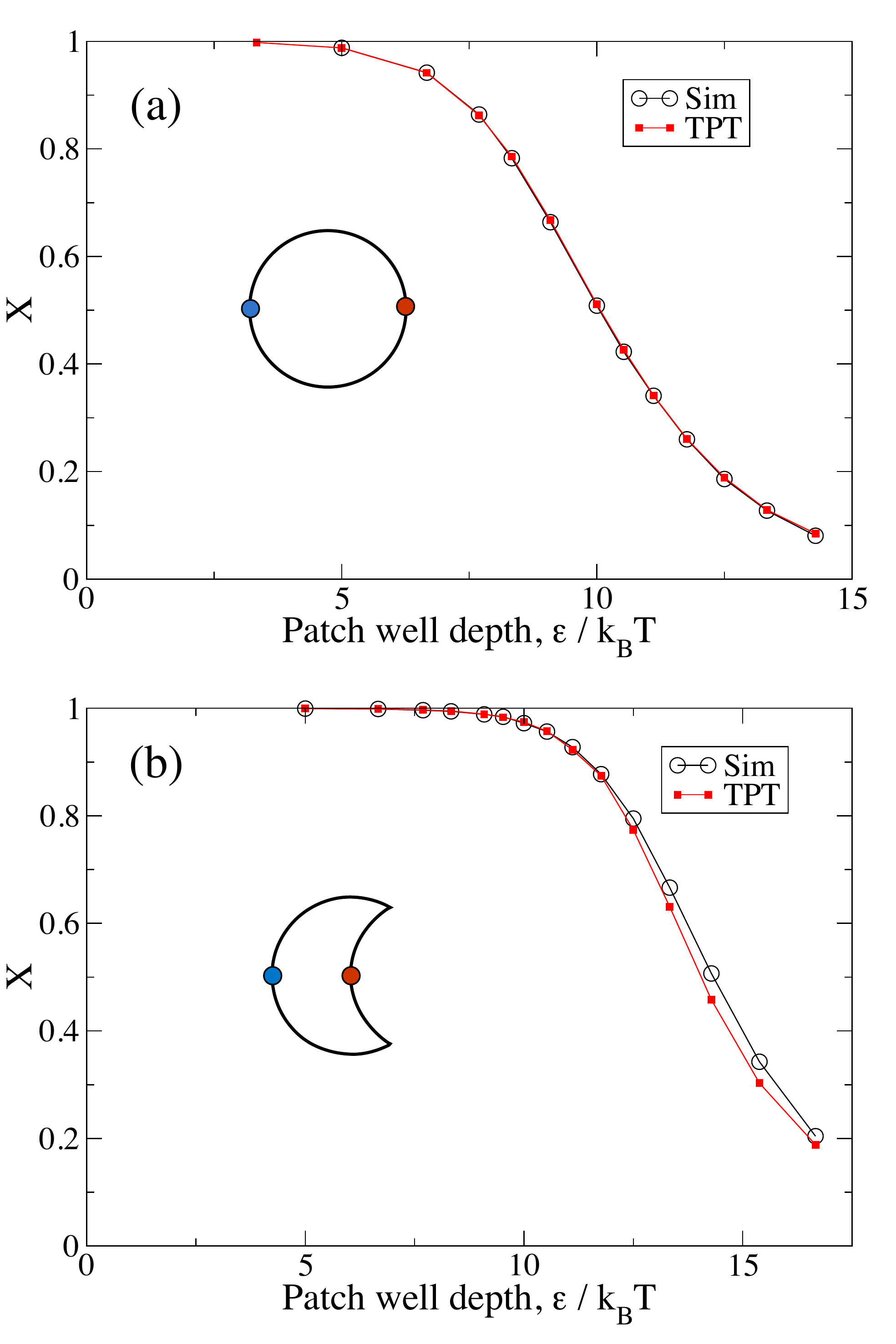}
  \caption{Comparison between simulation results and TPT predictions for the fraction, $X$, of unbonded $A$-patches in
  ``patchy'' models.  (a)~Hard sphere 
  model similar to that of Ref.~\cite{Sciortino2007}. 
  Each particle has an A patch and a B patch on opposite sides of the sphere, 
  the only interaction is between A and B and is a square well of range $\sigma_{AB}=0.119\sigma_l$ and strength (well-depth) $\varepsilon$.
  The agreement between theory and simulation is almost perfect.  
  (b)~Indented colloids ($h=0.5$) with patches.  This model system
  differs from the patchy spheres only in the particle shape and the patch location (there are no depletant particles).
  The A and B patches are located on the lock and key surfaces as shown, leading to lock-and-key binding.
  In this case, the non-spherical particle shape leads to significant deviations between simulation results and TPT predictions.
  Both panels show results at density $\rho=0.2\sigma_l^{-3}$.  
  The TPT results were obtained by numerical integration of (\ref{eq:TPT_K}),
  as in Fig.~\ref{fig:X}.
  }
  \label{fig:patchyspheres}
\end{figure}

For the spherical patchy particles considered by Sciortino and
co-workers \cite{Sciortino2007}, $g_{\rm R}$ can be approximated from
Percus-Yevick theory, and $f_{\rm A}$ is known exactly.  
Thus, the
TPT calculation can be performed analytically, and
it describes simulation data very accurately (see Fig.~\ref{fig:patchyspheres}a).
For indented colloids, neither $g_{\rm R}$ nor $f_{\rm A}$ is known
exactly, but $g_{\rm R}$ may be obtained from a simulation of
the reference system of indented colloids in the absence of depletant,
and $f_{\rm A}$ from a simulation of two particles in the presence of
depletant.  The integral in (\ref{eq:TPT_K}) can then be
calculated. 
We emphasise that the function $g_{\rm R}$ is evaluated in a system with no depletion interactions, so a single 
measurement of this function can be used (in conjunction with $f_{\rm A}$) to predict the behaviour for a wide range of $\eta$.
The resulting TPT predictions are shown in
Fig.~\ref{fig:X}(b) and (c) for $h=0.7$ and $h=0.5$ respectively: the agreement is reasonable but there are
deviations of up to $20\%$ between theoretical and simulation values for $X$.  
We attribute these deviations primarily to the differences in the packing properties 
of colloidal polymers, compared to isolated monomers.  As evidence for this, Fig.~\ref{fig:patchyspheres}b shows
results for indented colloids with ``patchy'' interactions.  This model system
differs from the patchy spheres only in the colloid shape and the patch location (there are no depletant particles)
-- it is clear that the TPT is less effective when colloids have non-spherical shapes. 
In the following paragraphs, we discuss how these shape (or packing) effects can be analysed within Wertheim's theory.  
 Another possible origin for deviations between theory and simulation 
in Fig.~\ref{fig:X} 
is that the TPT does not fully describe the attractive forces between colloids -- we discuss
this further at the end of this section.

\begin{figure} 
\includegraphics[width=8cm]{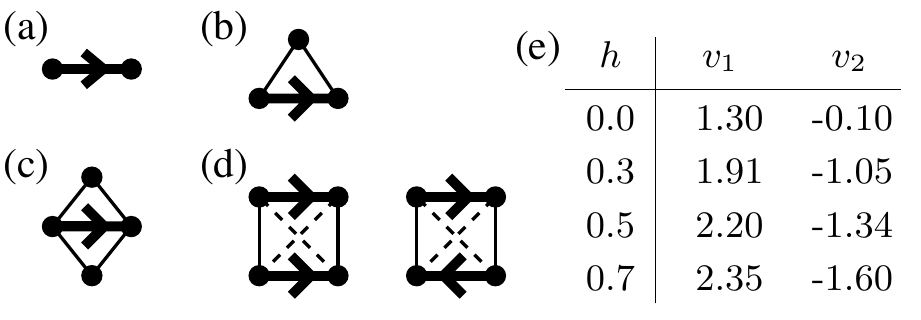}
  \caption{Liquid state theory diagrams showing contributions to the density functional that are
  relevant for calculating $\rhoLK$ within Wertheim theory. 
  Directed heavy lines correspond to attractive lock-and-key binding
  while thin lines correspond to repulsive interactions.  Where diagrams include
  dashed lines, this indicates a sum over diagrams both with and without these repulsive
  interactions.  Vertices have weights corresponding to various combinations of $\rho,\rhoL,\rhoK$,
  as prescribed by Wertheim's theory.
  (a)~Diagram for $K_0$; (b)~Diagram for the $v_1$-term in (\ref{eq:rob_K}); (c)~A diagram included
  in the TPT but not in SCW; (d) Diagrams for the $v_2$-term in Eq.~(\ref{eq:rob_K}), included in SCW theory but not in TPT.  
  The terms in (a,b,d) are those considered in the SCW theory.
  (e)~Table showing values
  of $v_1$ and $v_2$ for various $h$, in units having $\sigma_l=1$.}
  \label{fig:diagrams}
\end{figure}

To explore shape effects, we return to the diagrammatic analysis of Wertheim, but
instead of following the TPT, we consider
just a few simple terms in the density functional: see
Fig.~\ref{fig:diagrams}.  
Under this approximation, the law of mass
action is replaced by
\begin{equation}
\rhoLK = \rhoL \rhoK K_0 [ 1 + \rho v_1 + \rhoLK v_2 ]  .
\label{eq:rob_K}
\end{equation}
Here, $v_1$ and $v_2$ are geometrical factors (independent of $\vfs$) that account for packing
of free particles and short chains: the relevant liquid-state diagrams are shown in Fig.~\ref{fig:diagrams} while 
formulae for these quantities are given as supporting information~\cite{SI}.
Within Wertheim's TPT, the $v_1$-term is included, but the $v_2$-term is absent.  
Further, comparison between (\ref{eq:rob_K}) and the law of mass action shows that the effective equilibrium
constant $\rhoLK/(\rhoL\rhoK)$ now depends on the degree of polymerization of the system, via the $v_2$-term.  
This constant (and hence the quantity $X$) must therefore be determined
self-consistently by solving (\ref{eq:rob_K}), so we refer to the analysis in the presence of the $v_2$-term as self-consistent Wertheim (SCW) theory.

We have obtained values for $v_1$ and $v_2$ by simulating very small systems (up to $4$ 
indented particles, without depletant).  Results are shown in Fig.~\ref{fig:diagrams}(e).  Using these
values in (\ref{eq:rob_K}) leads to the SCW predictions shown in Fig.~\ref{fig:X}.  In terms of the density functional,
SCW theory 
is a much cruder approximation than TPT.  However,
the SCW theory performs significantly better than TPT: for $h=0.5$, it accounts for around half of the deviation between theory
and simulation.  

The  origin of this effect is the $v_2$-term in (\ref{eq:rob_K}). Physically, the $v_1$-term in that equation 
(included in both TPT and SCW theories)
reflects the increased virial pressure in the system as the colloid number
density increases, and enhances polymerization.  
The $v_2$-term reflects differences in packing properties between free particles and assembled chains.  
We find $v_2<0$, indicating that as polymerization occurs, the virial pressure is reduced (compared with TPT), suppressing further chain formation.
For spherical patchy particles ($h=0$), 
Fig.~\ref{fig:diagrams}(e) shows that the magnitude of $v_2$ is small, consistent with the success of TPT in that case (Fig.~\ref{fig:patchyspheres}).
However, the size of $v_2$ grows as $h$ is increased, leading to the deviations from TPT shown in Fig.~\ref{fig:X}.  The
SCW theory accounts for some of these deviations, although agreement is still not perfect.  Unlike the TPT, the 
SCW theory could be improved systematically, 
by including further terms in the density functional.  However, the theory presented here shows 
that packing effects of non-spherical particles can significantly affect self-assembly, and are required for quantitative
predictions.

Finally, we analyse one other possible origin for the deviations between TPT and simulation results
in Fig.~\ref{fig:X}.  As well as lock-and-key binding, the depletant particles also lead to an
attractive interaction between the concave (key) surfaces of the indented colloids.  This effect is
not accounted for in the TPT and SCW theories: we therefore considered a refined TPT in which
such ``key-key'' binding is included.  Let $\rho_{\rm KK}$ be the number density of key-key bonds 
(defined through a radial cutoff by analogy with $\rho_{\rm LK}$).  
Then the law of mass action predicts
$
 \rho_{\rm KK} = K_{\rm bb} \rho_{\rm K}^2
$
where $K_{\rm bb}$ 
is an equilibrium constant.  This key-key binding reduces the availability of particles for lock-and-key binding,
increasing the values of $X$ observed in simulation.  The TPT can be straightforwardly generalised
to include an estimate for $K_{\rm bb}$.  For those state points in Fig.~\ref{fig:X}(b,c)  
where the depletion interaction is strong, we find
 $K_{\rm bb} \ll K$, so lock-and-key binding is dominant; 
 when the interaction is relatively weak we find $\rho K_{\rm bb} \ll 1$, so bonds are rare.  Together, these results indicate
that key-key binding is a weak effect.  To verify this, we used $K$ and $K_{\rm bb}$ to calculate predictions
for $X$, noting
that the number density of key surfaces that are available for binding is 
$
  \rhoK = \rho  - \rhoLK - 2\rho_{\rm KK}.
$
As expected, the inclusion of the $\rho_{\rm KK}$ term in this equation reduces the TPT prediction for $\rhoLK$.  However,
the small value of $K_{\rm bb}$ means that this leads to a barely-discernible change in the TPT results
of Fig.~\ref{fig:X} (the differences between modified and unmodified TPT results are 
comparable with the line widths).  This reinforces our conclusion that it is primarily the shape and packing
properties of the indented colloids that lead to deviations from TPT predictions.

\section{Conclusions} \label{sec:conclusions}

The sophisticated Monte Carlo techniques that we have applied here
show that indented colloids can assemble into chains, with persistence
lengths and branching properties controlled by the colloid shape
and depletant density.  Two variants of Wertheim's theory have been analysed, showing
how particle shapes can influence their self-assembly.  
In addition to the colloidal polymers shown here, we also emphasize that the combination
of depletion interactions and colloidal shape has potential application for many other self-assembled
structures too, and that the simulation and theoretical methods used here can be readily applied in those cases.
Given that depletant parameters and the shapes of indented colloids can both be controlled in 
experiments~\cite{Sacanna:2010ys,Sacanna2011,sacanna13,Bahadur2012,Rossi11,Alsayed04,Savage09}, we hope that these results will
stimulate further experimental studies of self-assembly in these systems.

\footnotetext{\dag~Electronic Supplementary Information (ESI) available: [details of any supplementary information available should be included here]. See DOI: 10.1039/b000000x/}


\footnotetext{\textit{$^{a}$Department of Physics, University of Bath, Bath BA2 7AY,
  United Kingdom; Fax XX XXXX XXXX; Tel: XX XXXX XXXX; E-mail: d.ashton@bath.ac.uk}}


\footnotetext{\ddag~Additional footnotes to the title and authors can be included \emph{e.g.}\ `Present address:' or `These authors contributed equally to this work' as above using the symbols: \ddag, \textsection, and \P. Please place the appropriate symbol next to the author's name and include a \texttt{\textbackslash footnotetext} entry in the the correct place in the list.}


\footnotesize{
\bibliographystyle{rsc} 

}

\end{document}